\documentclass[useAMS, usenatbib]{mn2e}
\usepackage{subfigure}
\usepackage{ulem}
\usepackage{color}
\usepackage{graphicx}
\usepackage{times} 
\usepackage{amssymb}
\usepackage{amsmath}
\usepackage{lscape}
\usepackage{url}
\usepackage{multirow}
\usepackage{longtable}
\usepackage[english]{babel}
\usepackage{float}
\newif\ifAMStwofonts
\AMStwofontstrue

\newcommand{\mnras}{MNRAS}
\newcommand{\apj}{ApJ}
\newcommand{\aap}{A\&A}
\newcommand{\apjs}{ApJ Supplement}
\newcommand{\pasp}{PASP}
\newcommand{\apjl}{ApJ L}
\newcommand{\physrep}{Physics Reports}
\newcommand{\aaps}{A\&A Supplement}

\def\ave#1{\left\langle #1 \right\rangle}

\def\btheta{\vec{\theta}}
\title[Abell 2146 Weak lensing] {The Distribution of Dark and Luminous Matter in the Unique Galaxy Cluster Merger Abell 2146} 
\author[L. J. King et al.]
{\parbox[]{6.in}
{Lindsay\,J.\,King$^{1}$\thanks{E-mail: lindsay.king@utdallas.edu},\,Douglas\,I.\,Clowe$^{2}$, Joseph\,E.\,Coleman$^{1}$,\,Helen\,R.\,Russell$^{3}$, Rebecca\,Santana$^{2}$,\,Jacob\,A.\,White$^{1,4}$,\,Rebecca\,E.\,A.\,Canning$^{5,6}$,\,Nicole\,J.\,Deering$^{1}$, Andrew\,C.\,Fabian$^{3}$,\,Brandyn\,E.\,Lee$^{1}$,\,Baojiu\,Li$^{7}$,\,Brian\,R.\,McNamara$^{8}$ 
\\ } \\
\footnotesize
  $^{1}$Department of Physics, University of Texas at Dallas, 800 W. Campbell Road, Richardson, TX 75080, USA \\
  $^{2}$Department of Physics and Astronomy, Ohio University, 251B Clippinger Lab, Athens, OH 45701, USA\\
  $^{3}$Institute of Astronomy, University of Cambridge, Madingley Road, Cambridge CB3 0HA, UK\\
  $^{4}$Physics and Astronomy Department, University of British Columbia, 6224 Agricultural Road, Vancouver, BC V6T 1Z1, Canada\\
  $^{5}$Kavli Institute for Particle Astrophysics and Cosmology (KIPAC), Stanford University, 452 Lomita Mall, Stanford, CA 94305-4085, USA\\
  $^{6}$Department of Physics, Stanford University, 452 Lomita Mall, Stanford, CA 94305-4085, USA\\
  $^{7}$Department of Physics, Durham University, Durham, DH1 3LE, UK\\
  $^{8}$University of Waterloo, Department of Physics \& Astronomy, Waterloo, Canada
}
\begin{document}
\date{...}
\pagerange{\pageref{firstpage}--\pageref{lastpage}} \pubyear{....}
\maketitle
\begin{abstract}
Abell 2146 ($z$\,=\,0.232) consists of two galaxy clusters undergoing a major merger. The system was discovered in previous work, where two large shock fronts were detected using the {\it Chandra} X-ray Observatory, consistent with a merger close to the plane of the sky, caught soon after first core passage. A weak gravitational lensing analysis of the total gravitating mass in the system, using the distorted shapes of distant galaxies seen with ACS-WFC on {\it Hubble Space Telescope}, is presented. The highest peak in the reconstruction of the projected mass is centred on the Brightest Cluster Galaxy (BCG) in Abell 2146-A. The mass associated with Abell 2146-B is more extended. Bootstrapped noise mass reconstructions show the mass peak in Abell 2146-A to be consistently centred on the BCG. Previous work showed that BCG-A appears to lag behind an X-ray cool core; although the peak of the mass reconstruction is centred on the BCG, it is also consistent with the X-ray peak given the resolution of the weak lensing mass map. The best-fit mass model with two components centred on the BCGs yields $M_{200}$ = 1.1$^{+0.3}_{-0.4}$$\times$10$^{15}$\,M$_{\odot}$  and 3$^{+1}_{-2}$$\times$10$^{14}$\,M$_{\odot}$ for Abell 2146-A and Abell 2146-B respectively, assuming a mass concentration parameter of $c=3.5$ for each cluster. From the weak lensing analysis, Abell 2146-A is the primary halo component, and the origin of the apparent discrepancy with the X-ray analysis where Abell 2146-B is the primary halo is being assessed using simulations of the merger.
\end{abstract}

\section{Introduction}
Galaxy clusters are the most massive bound structures in the universe, forming at the intersections of filaments in the cosmic web and providing a sensitive test of the cosmological model and structure formation paradigm (e.g. Spergel \& Steinhardt 2000; Bahcall et al. 2003; Allen et al. 2011). Most of the mass in galaxy clusters is dark matter and the bulk of the baryonic mass is in the form of hot X-ray emitting plasma, comprising about 15\% of the total mass. Stars bound in cluster galaxies account for at most a few percent of the total cluster mass (e.g. Allen et al. 2011).

Massive galaxy clusters form from the hierarchical merger of groups and smaller clusters which collide at speeds of up to several thousand km\,s$^{-1}$. 
During a cluster merger, cluster galaxies behave like collisionless particles and are slowed only by tidal interactions. The hot plasma clouds behave in a different manner and slow down as they pass through each other, since they are affected by ram pressure. Shortly after each collision in the merger process, the plasma clouds are expected to lag behind the major concentrations of cluster galaxies, for example as seen in 1E0657-56, the ``Bullet Cluster" (Markevitch et al. 2004; Clowe et al. 2004; Brada\v{c} et al. 2006; Clowe et al. 2006). Dark matter is expected to be located near to the cluster galaxies, since it does not have a large self-interaction cross-section (e.g. Randall et al. 2008). Thus the dominant baryonic component can be offset from the bulk of the total mass. Clusters that have recently undergone a major merger close to the plane of the sky are very rare systems, but they are extremely important events: as well as investigating the properties of dark matter, these systems are very promising laboratories for the study of the hot plasma in clusters and the physical transport processes in the intracluster medium (ICM) (e.g. Russell et al. 2012). They can also be used to test the $\Lambda$CDM paradigm, and alternative theories of gravity and models for dark energy, through for example their pair-wise velocity distribution (e.g. see the review in Clifton et al. 2012). Major mergers between two massive clusters are the most energetic events since the Big Bang. The kinetic energy of the systems can reach $\sim$10$^{57}$\,J, and a significant fraction is dissipated by such large-scale shocks driven into the ICM and by subsequent turbulence in the post-shock regions and ICM (e.g. Sarazin 2001; Markevitch \& Vikhlinin 2007). 

In addition to the Bullet Cluster, several other merging cluster systems have been studied in detail, for example MACS J0025.4-1222 \citep{BR08.1}, Abell 1758 \citep{OK08.1,RA12.1}, Abell 754, Abell1750, Abell 1914, Abell 2034, Abell 2142 \citep{OK08.1}, Abell 2744 \citep{ME11.1}, Abell 2163 (Okabe et al. 2011; Soucail 2012),  and Abell 520 (e.g. Clowe et al. 2012). Some of these systems are complex however, involving several primary clusters undergoing mergers, or with merger axes with a large angle to the plane of the sky, making analysis and interpretation more challenging. 

\begin{figure*}
\begin{center}
\includegraphics[width=18cm]{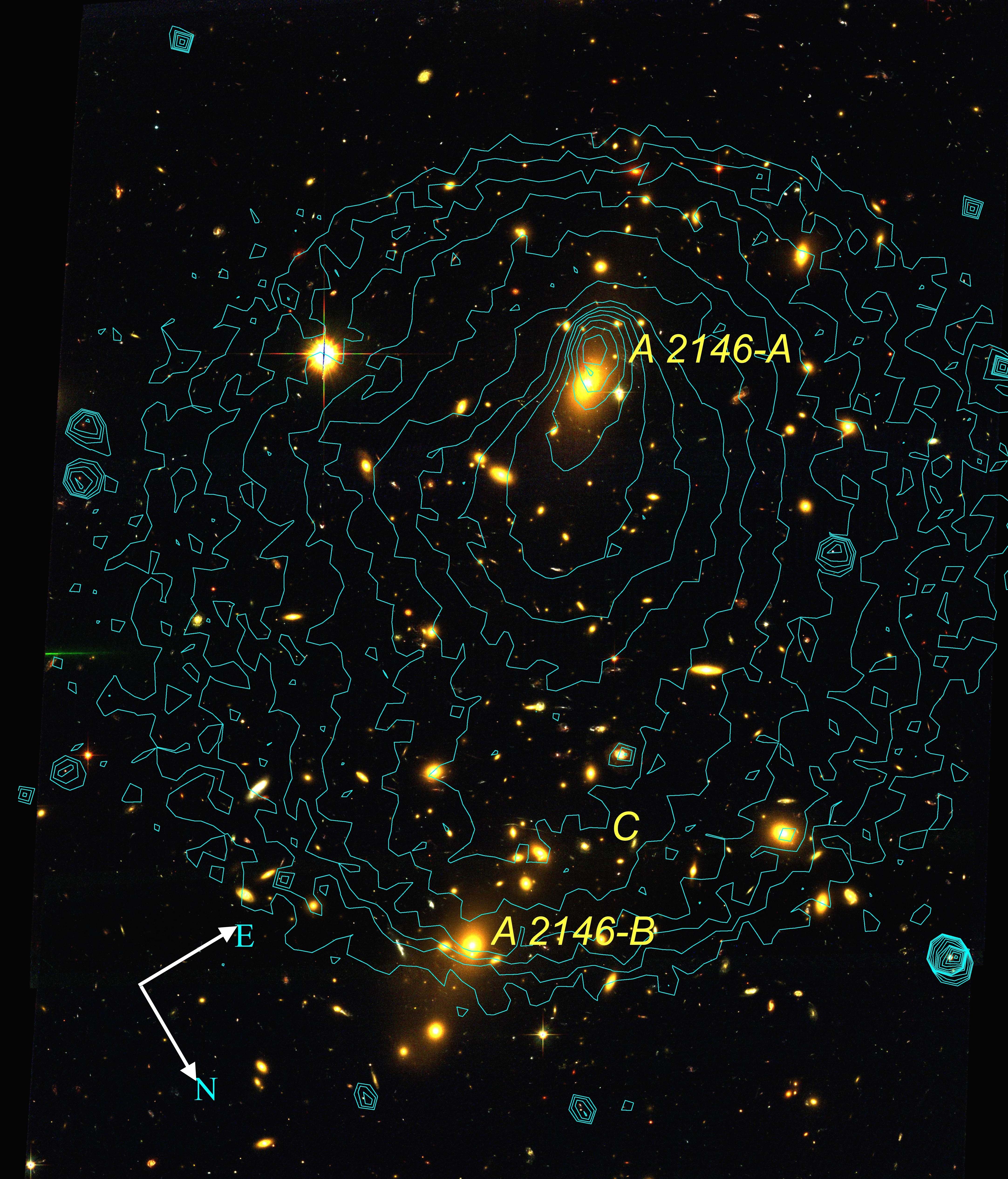}
\caption{Colour composite of Abell 2146 from {\it HST} F435W, F606W and F814W observations. Labels for Cluster Abell 2146-A and cluster Abell 2146-B are placed to the east of their Brightest Cluster Galaxies. The label C to the east of the BCG in Abell 2146-B is discussed in Section 4. Contours show the X-ray intensity from {\it Chandra} X-ray Observatory as described in Russell et al. (2012). Note that in this figure East is to the top-right and North is to the bottom-right as indicated, and the X-ray contours from Russell et al. (2012) are rotated accordingly for comparison with the {\it HST} composite.}
\label{hst}
\end{center}
\end{figure*}

The nature of Abell 2146 (Struble \& Rood 1999) as a merger system was first realised by Russell et al. (2010) who mapped the hot gas structure using the {\it Chandra} X-ray Observatory. These observations revealed an X-ray morphology similar to that of the Bullet Cluster, consistent with two massive galaxy clusters having undergone a recent merger with first core passage $\approx 0.1-0.3$\,Gyr ago, and still moving away from each other. The existence of 2 large shock fronts (Mach number $M\sim 2$) is unique among these merger systems, and is indicative of clusters which are closer in mass than those in the Bullet Cluster system (e.g. Markevitch et al. 2004; Mastropietro \& Burkert 2008; Lage \& Farrar 2014).  Deeper {\it Chandra} observations of the system are presented in Russell et al. (2012). 

We refer to what appears to be the ``bullet" cluster component on X-ray maps of Abell 2146 as Abell 2146-A, and to the other cluster component as Abell 2146-B. It has been established that the location of the X-ray cool core of Abell 2146-A is offset from the location of the BCG by 36\,kpc. However, remarkably, the cool core {\it leads} rather than {\it lags} the BCG. In Abell 2146-B, the centroid of the galaxies is leading the bulk of the plasma, as expected, with the shock front being almost coincident with the BCG. The origin of the direction of the offset in Abell 2146-A is unclear, possibly being due to perturbation by another galaxy, or to a merger that is somewhat off axis (Canning et al. 2012; White et al. 2015). BCGs are very rarely seen to lag behind the ICM in merger systems. In Abell 168, Hallman \& Markevitch (2004) suggested that the BCG lagging the ICM is due to a ``ram pressure slingshot", resulting from a drop in ram pressure on the plasma when the sub-cluster is approaching the apocentre of its orbit, at a late stage in the merger. In the complex merger system Abell 2744, the ICM also leads the galaxies and dark matter in one of the four clusters undergoing a merger; Owers et al. (2011) and Merten et al. (2011), however, suggest that a ram pressure slingshot is responsible. 
The direction of the offset of the eastern mass and X-ray peaks in Abell 754 are also indicative of the eastern mass component reaching the apocentre of the merger orbit, and falling back towards the centre for the second core passage (Okabe \& Umetsu 2008). This effect would not be expected in Abell 2146, since it is observed at an earlier stage in the merger; Russell et al. (2010) estimated that the time scale for a ram pressure slingshot to occur would be $\approx$\,1\,Gyr after first core passage, several times longer than the age estimated from observations. 

Canning et al. (2012) estimated that in Abell 2146 the merger axis is inclined at only $\sim17^{\circ}$ to the plane of the sky, using the line of sight velocity difference between the brightest cluster galaxies in each of the clusters along with the X-ray shock velocities. A dynamical analysis of cluster galaxies presented in White et al. (2015) is also consistent with a recent merger that is relatively close to the plane of the sky, with a merger axis inclined at $13^{\circ}-19^{\circ}$ and a time scale since first core passage of $\approx 0.12-0.14$\,Gyr. In addition, the detection of the shock fronts with {\it Chandra} in itself requires a relatively recent merger with a small angle to the line of sight; a larger angle would result in smearing of the sharp surface brightness edges when seen in projection, and in a system observed later in the merger process the shock fronts would have travelled further into the low density region and would go undetected. 

Mass estimates for the system have been obtained using several different techniques. Using a mass - X-ray temperature scaling relation (e.g. from Finoguenov et al. 2001) yields a mass estimate of $M_{\rm 500} \sim7 \times 10^{14} M_{\odot}$ \footnote{Throughout we use $M_{n}$ to denote the mass inside the radius $r_{n}$ where the mean mass density is $n$ times the critical density at the redshift of halo formation.}, with X-ray observations indicating that Abell 2146-A is the lower mass cluster (Russell et al. 2010). The system has also been detected in Sunyaev-Zel'dovich (SZ) observations; clusters distort the intensity of the Cosmic Microwave Background (CMB) when about 1\% of CMB photons undergo inverse-Compton scattering and gain energy from the electrons in the intracluster gas (Sunyaev \& Zel'dovich 1970). The SZ signal was measured using the Arcminute MicroKelvin Imager (AMI), with a peak signal-to-noise ratio of 13$\sigma$ in the radio source subtracted map (AMI Consortium: Rodr\'iguez-Gonz\'alvez et al. 2011). The total mass inside $r_{200}$ estimated by the AMI Consortium from the SZ signal is $4.1\pm 0.5 \times 10^{14} h^{-1}M_{\odot}$. The total dynamical mass estimated by applying the virial theorem to spectroscopic observations of cluster members in the system is $M_{\rm vir}= 8.5^{+4.3}_{-4.7}\times 10^{14} M_{\odot}$ (White et al. 2015), not corrected downwards for a surface pressure term of $\approx 20\%$ (The \& White 1986), with Abell 2146-A being the higher mass cluster. 

Deep radio observations of Abell 2146 by Russell et al. (2011) using the Giant Metrewave Radio Telescope (GMRT) at 325\,MHz do not detect an extended radio halo or radio relics associated with the shock fronts, at odds with all other merging galaxy clusters with X-ray detected shock fronts, including the Bullet Cluster, Abell 520 and Abell 754, and with candidate shock fronts. The radio power expected from the P$_{\rm{radio}}$-L$_{\rm X-ray}$ correlation for merging systems of Cassano et al. (2013) is significantly higher than the measured upper limit, which remains a puzzle. However, see the discussion in White et al. (2015) of the absence of a detected radio halo in Abell 2146 in the context of the P$_{\rm{radio}}$-$M_{500}$ correlation of Cassano et al. (2013).

Gravitational lensing is sensitive to the total gravitating mass of the system, probing both dark and luminous matter. In this paper we present a weak gravitational lensing analysis of Abell 2146, using the distorted shapes of distant galaxies on ACS-WFC {\it Hubble Space Telescope} images (PI: King, proposal 12871). In Section 2 we describe the {\it HST} observations. In Section 3 we outline the relevant aspects of weak lensing and describe how the catalogues of galaxies used in the weak lensing analysis were obtained. We present weak lensing mass maps and parameterized mass models of the system in Section 4. The results are discussed in Section 5, and we conclude in Section 6.

Throughout this paper, for comparison with previous work, we assume a $\Lambda$CDM cosmology with present day Hubble parameter $H_{0}$\,=\,$70\,{\rm km}\,{\rm s}^{-1}\,{\rm Mpc}^{-1}$, and present day matter density and dark energy density parameters $\Omega_{\rm M}$\,=\,0.3 and $\Omega_{\Lambda}$\,=\,0.7 respectively. We assume dark energy to be a cosmological constant, equation of state parameter $w=-1$. At the redshift of Abell 2146 ($z=0.2323$), the physical scale is 3.702 kpc per arcsecond, and the Hubble parameter $H$\,=\,$78.7\,{\rm km}\,{\rm s}^{-1}\,{\rm Mpc}^{-1}$.

\section{Hubble Space Telescope Observations}
We obtained 8 orbits of {\it HST} optical imaging with ACS/WFC on 2013 June 3 and June 6 ({\it HST} Cycle 20 proposal 12871, PI: King). The data consist of two pointings in each of the F435W and F606W filters, and four pointings in the F814W filter. The corresponding exposure times are 5344\,s in F435W, 5360\,s in F606W and 10552\,s in F814W. Each orbit was split into 4 dither positions, with a small offset between the first two images, a chip gap spanning offset before the 3rd image, and another small offset for the fourth image.                                                                

Since the primary goal of the programme was weak lensing analysis, requiring that the ellipticities of galaxies should be measured as accurately as possible, special care was taken when reducing and combining the images. We followed a similar procedure to that described in Clowe et al. (2012).

Due to being above the protection of the Earth's atmosphere, energetic particles have damaged the CCD detectors of ACS, creating ``hot" pixels and charge traps. 
After each exposure, during the transfer of photoelectrons through the silicon substrate to the readout electronics, a fraction is temporarily retained by lattice defects 
and released after a short delay \citep{JA01.1}. This so-called ``Charge Transfer Inefficiency" (CTI) effect spuriously elongates the observed shapes of galaxies, in particular faint galaxies, in a way that 
mimics weak gravitational lensing. In addition, all images taken with ACS/WFC after Servicing Mission 4 show a row-correlated noise (striping) due to the CCD Electronics Box Replacement. 

We used the debiased, CTI-corrected, striping-corrected and flat fielded images provided by STScI. These were used as the input to a modified version of the HAGGLeS pipeline, provided by Tim Schrabback (Schrabback 2008), to do the distortion corrections and determine the image alignments.  The Multidrizzle algorithm \citep{multidrizzle} was used to do the final coaddition with the alignments determined by HAGGLeS. 

Magnitudes were corrected for Milky Way dust extinction using the Schlafly \& Finkbeiner (2011) estimates from an analysis of Sloan Digital Sky Survey data, which prefer a Fitzpatrick (1999) reddening law. The corrections require subtracting 0.108, 0.074 and 0.045 from the magnitudes in F435W, F606W and F814W respectively. 
The number counts of the final resulting images depart from an exponential growth function at $m_{\mathrm{F435W}} = 26.2$, $m_{\mathrm{F606W}} = 25.9$, and $m_{\mathrm{F814W}} = 25.5$ for each of the 3 filters, with photometry measured using {\textsc Source Extractor} (Bertin \& Arnouts 1996).  A colour composite of the central region of the observations is shown in Fig.\,\ref{hst}, with superposed contours of X-ray intensity from deep {\it Chandra} observations (Russell et al. 2012).

\section{Weak Gravitational Lensing}
In this section we describe how we obtained the catalogues of galaxies used in the weak lensing analysis. First, we summarise the key aspects of gravitational lensing relevant to our analysis; see for example Bartelmann \& Schneider (2001) for a detailed review. 
\
The gravitational lensing potential, $\Psi$,  is a scaled projection of the Newtonian gravitational potential $\Phi$:
\begin{equation}
\Psi(\btheta) =  \frac{2}{c^2}\frac{D_{\rm LS}}{D_{\rm L}D_{\rm S}}\int {\rm d}z\,\Phi(D_{\rm L}\btheta, z)\,,
\end{equation}
where angular position in the lens plane is denoted by $\btheta$, $D_{\rm L}$, $D_{\rm S}$ and $D_{\rm LS}$ are the lens, source and lens-source angular diameter distances respectively, and the integral is over redshift $z$. 

A cluster system has an extent much less than any of these cosmological distances, so that the thin-lens approximation holds. For a discrete mass with two-dimensional projected surface mass density $\Sigma(\btheta)$, the dimensionless lensing convergence is defined as: 
\begin{equation}
\kappa(\vec\theta) = \frac{\Sigma(\vec\theta)}{\Sigma_{\rm crit}},~~{\rm where}~~\Sigma_{\rm crit} = \frac{c^{2}}{4\pi G}\frac {D_{\rm S}}{D_{\rm L}D_{LS}}\,,
\end{equation}
where we introduced the critical surface mass density $\Sigma_{\rm crit}$; a sufficient condition for multiple image production by a lens is that $\kappa\,\textgreater\,1$, i.e. $\Sigma\,\textgreater\,\Sigma_{\rm crit}$.
We also introduce the shear due to the lens, a spin-2 quantity that can be compactly written as a complex number, $\gamma(\vec\theta)=\gamma_{1}+i\gamma_{2}$. With $\gamma =|\gamma| e^{2i\alpha}$, the strength and position angle $\alpha$ of the shear are given by the modulus and half of the phase of the complex number respectively.

We write both $\kappa$ and $\gamma$ quantities in terms of linear combinations of second derivatives of the lensing potential:
\begin{equation}
\kappa = \frac{1}{2}\nabla^{2}\Psi\,;
~~\gamma_{1} = \frac{1}{2}\left(\frac{\partial^{2}\Psi}{\partial\theta_{1}^{2}}-\frac{\partial^{2}\Psi}{\partial\theta_{2}^{2}}\right);
~~\gamma_{2} = \frac{\partial^{2}\Psi}{\partial\theta_{1}\partial\theta_{2}}\,.
\end{equation}
The convergence by itself results in an isotropic focusing of a light bundle, and the shear induces anisotropic distortions in the observed shape. For a source that is much smaller than the scale on which the properties of the lens change, and denoting angular position in the source plane by $\vec\beta$, the image distortion is given by a Jacobian matrix:
\begin{equation}
{\cal A(\vec\theta)} \equiv \frac{\partial\vec\beta}{\partial\vec\theta} = 
\left( 
\begin{array}{cc}
	1 -\kappa -\gamma_{1}  & -\gamma_{2} \\
	-\gamma_{2}  & 1 -\kappa +\gamma_{1}\\
	\end{array}\right)\,,
\end{equation}
and it can be seen that $\kappa$ appears only in diagonal elements, consistent with isotropic distortion. In the regime where $\kappa\ll 1$ and $\gamma\ll 1$, weak lensing results in single, distorted images of distant galaxies which are only slightly magnified; taking the inverse of the determinant of the Jacobian matrix yields a magnification $\mu \approx 1 + 2\kappa$ in the weak lensing regime. 

Analogous to the shear due to the lens, the shape and position angle of a source galaxy can be described by a complex ellipticity $\epsilon^{\rm s}$, with modulus $|\epsilon^{\rm s}| = (1-b/a)/(1+b/a)$, where $b/a$ is the ratio of the minor to major axis. The phase of $\epsilon^{\rm s}$ is twice the position angle of the galaxy $\phi$, 
$\epsilon^{\rm s}=|\epsilon^{\rm s}| e^{2i\phi}$. In the weak lensing regime, the complex ellipticity of a lensed galaxy $\epsilon$ is then given by
\begin{equation}
\epsilon = \frac{\epsilon^{\rm s} + g}{1+g^{*}\epsilon^{\rm s}}\approx \epsilon^{\rm s} + \gamma
\end{equation}
where $g=\gamma/(1-\kappa)$ is the complex reduced shear and * denotes complex conjugation. The observed ellipticity of a galaxy is a noisy measure of the shear since galaxies have an intrinsic distribution of shapes and orientations. Of great importance in weak lensing is that the expectation value of the lensed ellipticity over a small patch of sky is $\ave{\epsilon}=g\approx \gamma$. Thus after accounting for ellipticity distortions due to imperfect optics, and the atmosphere for ground-based observations, we can use the ellipticities of galaxies to estimate the shear field, and hence reconstruct the projected mass based on relationships in real or Fourier space (e.g. Kaiser \& Squires 1993; Seitz \& Schneider 2001). We can also fit parameterized mass models to the ellipticities (e.g. Schneider, King \& Erben 2000) to obtain the best fit mass density distributions.  

\subsection{Obtaining galaxy catalogues}
The goal of the weak lensing analysis is to obtain a map of the convergence $\kappa$ (or surface mass density $\Sigma$) of the system, and also to fit parametric mass models to the cluster components. In order to achieve this, we first need to estimate the shear field from the shapes and orientations of distant galaxies.
   
Since telescope optics are not perfect, the measured ellipticity of a distant galaxy must be corrected for Point Spread Function (PSF) smearing to give a measurement of the shear. The methodology that we use for PSF correction is a modified KSB technique (Kaiser, Squires \& Broadhurst 1995), with modifications as described in Clowe et al. (2006b), using stars on the images to determine the corrections. 

Weak lensing measurements were performed with a modified version of the {\textsc IMCAT} software package created by Nick Kaiser. A combined shear catalogue was obtained from weak lensing analysis performed separately on each of the ACS pointings.

\begin{figure}
\begin{center}
\includegraphics[width=9cm]{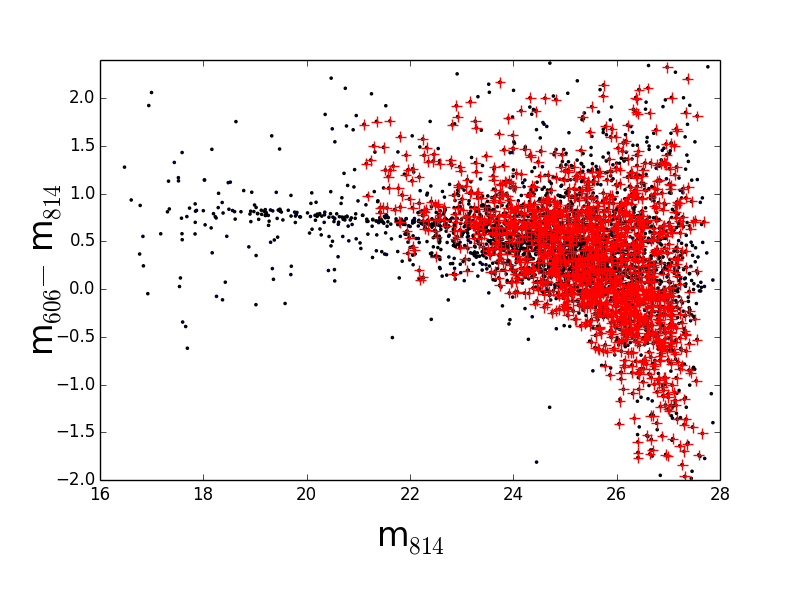}
\caption{Colour-magnitude diagram showing objects in the field of Abell 2146. Note the well defined line of dense points corresponding to the red sequence of cluster galaxies. Those objects selected as background galaxies using the criteria in the text are indicated by red +.}
\label{crag}
\end{center}
\end{figure}

\begin{figure}
\begin{center}
\includegraphics[width=9cm]{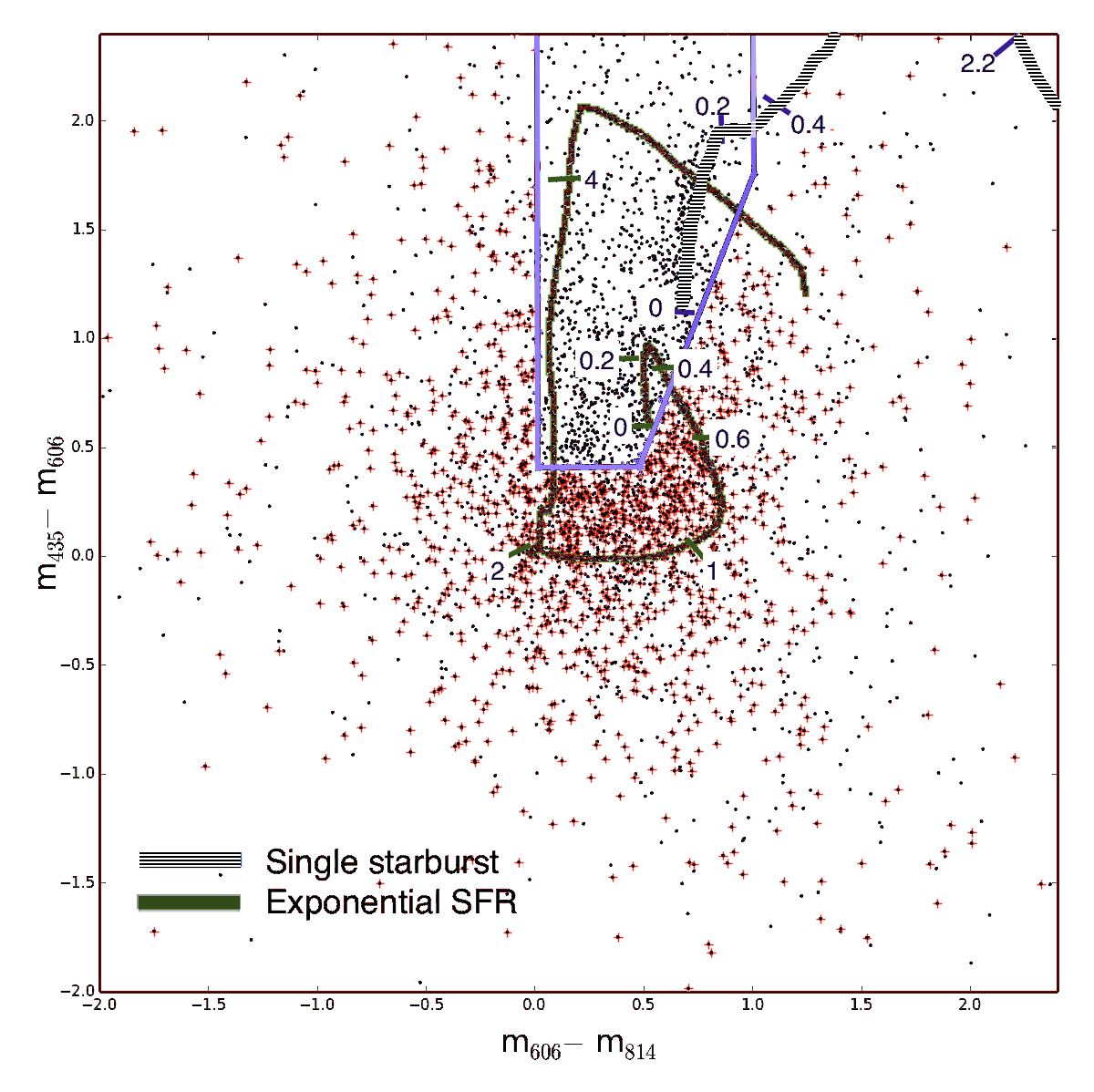}
\caption{Colour-colour diagram showing objects in the field of Abell 2146 (small black circles), and those selected as background galaxies using the criteria in the text (red +). Objects inside the region bounded by the solid lavender lines are excluded by the colour cuts designed to select background galaxies as described 
in the text. Note the line of dense points inside the excluded region corresponding to the red sequence of cluster galaxies.The dashed line and green line show
theoretical colours for galaxies with a present day solar metallicity, formed at $z = 6$ with a Salpeter initial mass function (Salpeter 1955) and with
a single starburst population (dashed line) or a 10 Gyr exponential decay star formation rate (green line). The tick marks and labels on the lines
indicate where galaxies of a given redshift are expected to reside on the colour-colour diagram. The models for the evolution of colours were generated using the EzGal 
software (Mancone \& Gonzalez 2012) with an updated Bruzual \& Charlot (2003) model.}
\label{crag2}
\end{center}
\end{figure}

Stars were identified based on a size cut ($<$0.081$\arcsec$ for $50\%$ encircled light radius) and objects with an unusually high central surface brightness for their magnitude were also rejected.  The KSB PSF correction terms were measured for a range of weighting function sizes, and were fit using a 3rd order polynomial for image position variations in each pointing.  The fitted values matched to the galaxy size were used to correct for the PSF smearing.  The PSF corrections were calibrated using the ACS-like STEP3\footnote{(http://www.roe.ac.uk/$\sim$heymans/step/cosmic\_shear\_test.html)} simulations, accounting for a systematic underestimate of $\sim$$8\%$ in the shear measurements.  We performed the shear measurements independently for each of the three ACS passbands. 

Weights for each galaxy in each data set were obtained by computing the inverse of the rms shear for nearby neighbours in significance and size space, with each data set showing that large, bright galaxies have a rms intrinsic shape of $\sigma_g = 0.245$ per shear component in the F814W passband, $0.265$ for the F606W passband, and $0.27$ for the F435W passband. The increasing measurement errors for the second moments of fainter and smaller galaxies lead to them having larger rms shear values than these. The shear estimates were combined across the different passbands following the procedure described in Clowe et al. (2012).

Weak lensing analysis requires that we select galaxies that are more distant than the lens. Likely cluster members and foreground objects were rejected based on their brightness and colour. The weak lensing analysis was restricted to faint galaxies with a {\textsc Source Extractor} AUTO magnitude $>$ 21, signal-to-noise S/N $>$ 5 in the F814W passband and with no bright neighboring galaxies that would impact on the measurement of their brightness. Further, the selection used here includes galaxies with S/N $>$ 10 in at least 1 passband, and with photometric measurements in all 3 passbands.

The colour cuts were based on templates from the EzGal software package (Mancone \& Gonzalez 2012), specifically using an updated Bruzual \& Charlot (2003) model to assess the redshift evolution of the apparent magnitudes of stellar populations viewed through the ACS filters. Stars with present day solar metallicity formed at $z = 6$ with a Salpeter initial mass function (Salpeter 1955), and with either a single starburst population or a 10 Gyr exponential decay star formation rate. Knowing where galaxies of a given redshift are expected to reside on the colour-colour diagram guides colour cuts to exclude potential cluster members or foreground objects within the region defined by $0 < {\rm m}_{606} - {\rm m}_{814} < 1$; $0.4 < {\rm m}_{435} - {\rm m}_{606} < 3$;  ${\rm m}_{435} - {\rm m}_{606} > 2.6 ({\rm m}_{606} - {\rm m}_{814}) - 0.9$, where the cuts hold simultaneously.  This results in a set of 1520 galaxies that we focus on in this paper, corresponding to a number density of 76 galaxies per arcmin$^{2}$.

Fig.\,\ref{crag} shows a colour-magnitude diagram for all of the objects detected in the field and for the galaxies used in the lensing analysis; note the well defined cluster red sequence. Fig.\,\ref{crag2} shows a corresponding colour-colour diagram. The region bounded by solid lavender lines is excluded by the colour cuts noted above. The red sequence of cluster galaxies can be seen as an over-density inside this excluded region. The dashed line and the green line show the evolution with redshift of the theoretical colours for stellar populations as described above.

\section{Results}
Reconstructions of the projected mass distribution were carried out using the Seitz \& Schneider (1995) modification of the Kaiser \& Squires (1993) technique. This method uses the observed complex ellipticities of galaxies to estimate the shear field and hence to map the convergence field. 

\begin{figure}
\begin{center}
\includegraphics[width=8cm]{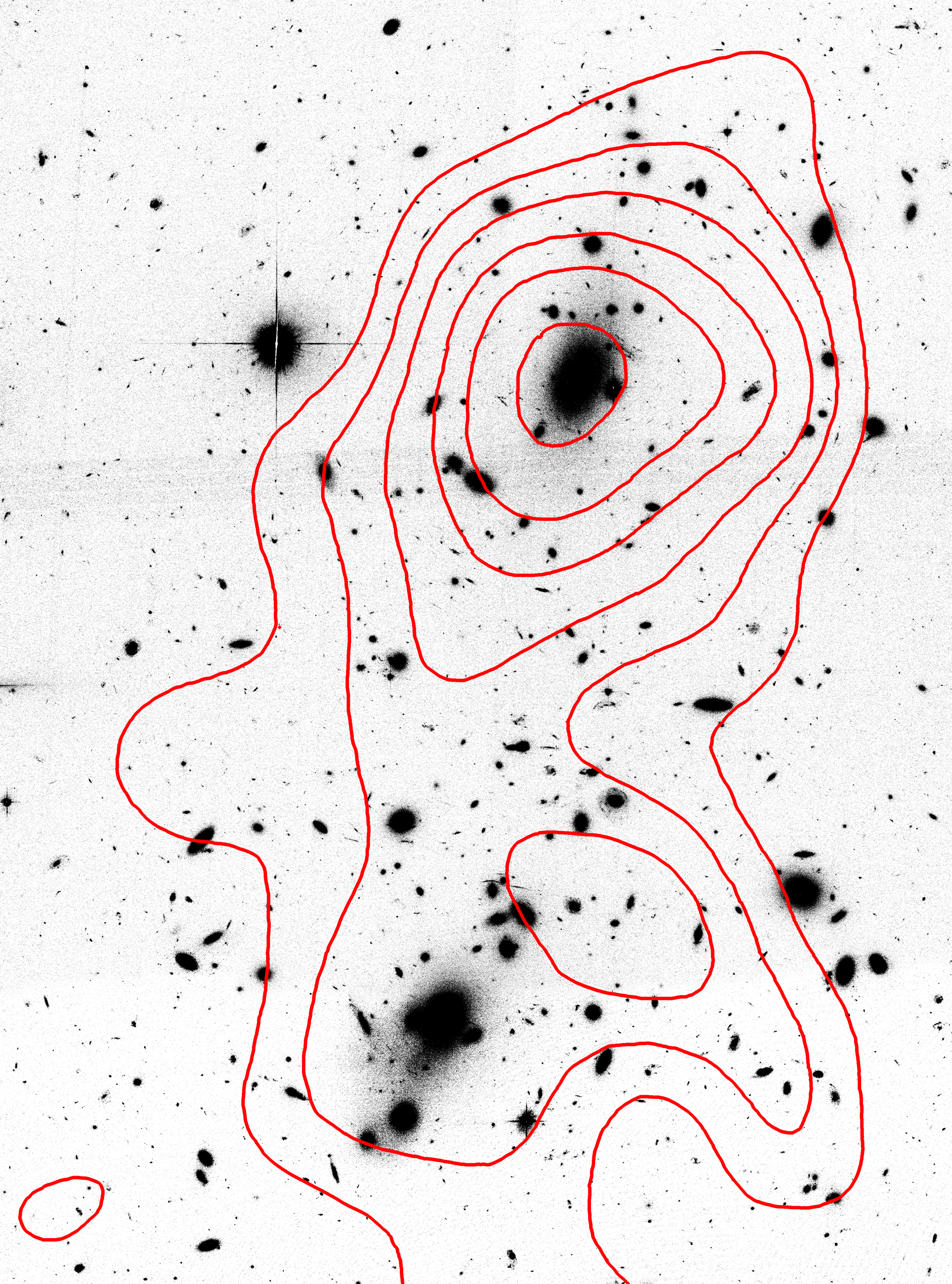}
\caption{Mass reconstruction of the Abell 2146 system obtained using the selection of background galaxies described in the text. Contours of convergence are plotted starting at $\kappa = 0.13$ and increasing inwards in steps of 0.03. The background greyscale is the {\it HST} F814W image.}
\label{diff}
\end{center}
\end{figure}

\begin{figure}
\begin{center}
\includegraphics[width=6cm]{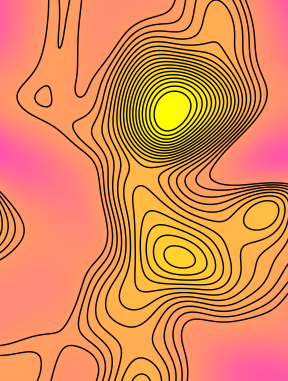}
\end{center}
\begin{center}
\includegraphics[width=0.5cm]{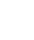}
\end{center}
\begin{center}
\includegraphics[width=6cm]{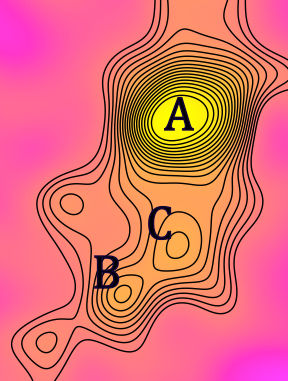}
\end{center}
\caption{Two representative weak lensing convergence reconstructions from the 30,000 bootstrap resampled catalogues. In bootstrap resampling, new catalogues with the same number of entries are obtained by selecting galaxies at random (with replacement) from the original galaxy catalogue. Contour levels are plotted at smaller linear intervals than in Fig.\,\ref{diff} to illustrate the behavior of the mass reconstruction of A2146-B with the bootstrap resampled catalogues. The mass peak in A2146-A is consistently coincident with the BCG in the cluster, whereas the location of the mass peak in A2146-B varies with resampled catalogue. This is quantified in Fig.\,6.}
\label{diffex}
\end{figure}

The selected background galaxies were used to reconstruct the 2-D convergence, as shown in Fig.\,\ref{diff}. The smoothing scale for the mass maps, determined by the number density of galaxies and their intrinsic ellipticity dispersion, is 14.4 arcsec ($\approx 50$\,kpc at the system redshift). Since we do not have observations in enough filters to estimate photometric redshifts for the objects in the field, we take the background galaxies to be at $z_{s}=0.8$. This source redshift is used in calculating the critical surface mass density, and hence the surface mass density of the lens. Changing this to $z_{s}=0.85$ as in Clowe et al. (2012) would decrease the reported masses systematically by $\sim$2.5\%. 

Determining the errors on a mass map is rather complex, since variations in the number density of background galaxies and in their intrinsic ellipticity cause the errors to vary by a large amount over the reconstructed map.  A method that is often used to estimate the error on a mass map is to measure the rms shear and the mean number density of the background galaxies, and to obtain an average noise level based on propagating these errors through the mass reconstruction algorithm. However, this approach of average noise level determination can be problematic, since the measurement of the convergence around a given peak is derived from the shear of background galaxies with an effective weighting of $\gamma/r$, where $r$ is the projected distance of a background galaxy from the peak. This means that most of the weight comes from the ellipticities of a small number of background galaxies close to the peak, where $\gamma$ is largest and $r$ is smallest. If any of these galaxies has an extreme intrinsic ellipticity, the noise in that peak will be much larger than average. 

The noise in mass reconstruction stems primarily from the distribution of the intrinsic ellipticities (shapes and orientations) of background galaxies. In order to preserve the underlying reduced shear field while assessing the impact of noise, we use bootstrap resampling of the catalogue of galaxies used in the mass reconstruction. In this process, galaxies are randomly selected (preserving their positions and complex ellipticities) from the original catalogue, with replacement, to obtain a new catalogue with the same number of entries as the original. This means that a particular galaxy can appear in the new catalogue more than once (i.e. at the same location and with the same complex ellipticity), or not appear at all. For a large catalogue, the chance of any given galaxy having an integer weight $W\ge 0$ is $e^{-1}/W!$ - e.g. the probability that the same galaxy appears twice is $1/2e$ and so on - and the probability of any group of $N$ galaxies not being in the new catalogue is $e^{-N}$. Examples of mass reconstructions made from two different bootstrapped catalogues are shown in Fig.\,\ref{diffex}. 

\begin{figure}
\begin{center}
\includegraphics[width=8cm]{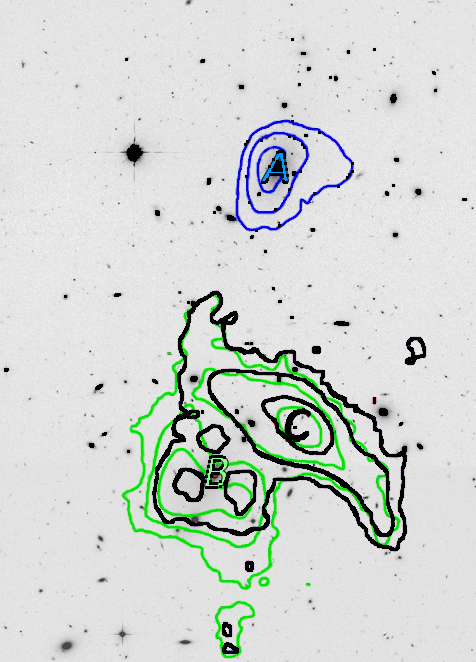}
\caption{For 30,000 mass reconstructions made using bootstrapped resampled catalogues of the selected background galaxies, contours enclosing 68.27\%, 95.45\%, and 99.73\% of mass peaks nearest to fixed points A (centred on the BCG in Abell 2146-A), B (centred on the BCG in Abell 2146-B), and C (a point where many of the bootstrapped mass maps show a high peak) are shown in blue, green, and black respectively. Around A, all of the most significant peaks are found close to the BCG. In about half of the mass maps, the peak closest to B is found closer to C, and about 25\% of the peaks closest to C are found closer to B. On about 1\% of mass maps there is no significant peak near to either B or C, instead the closest peak is near to A or is a noise peak.}
\label{boot}
\end{center}
\end{figure}

We now focus on discussing the errors on the mass reconstruction. We bootstrap resampled the selected background galaxies, with replacement, and created 30,000 new catalogues with the same number of entries as the original.  Next, we made 2-D mass maps using each of the bootstrap resampled catalogues, and determined statistics from these maps as follows. Taking as fixed points the BCG in Abell 2146-A (fixed point A), the BCG in Abell 2146-B (fixed point B) and the centre of a mass peak located just east of Abell 2146-B (fixed point C), the location of the nearest significant mass peak is determined on each of the bootstrap mass maps. In Fig.\,\ref{boot} contours that enclose 68.27\%, 95.45\%, and 99.73\% of the mass peaks are shown for fixed points A, B and C in blue, green and black respectively. In all of the reconstructions, the closest significant peak to point A is located near to point A. In about half of the reconstructions, the closest peak to point B is closer to point C. When centred on point C, about 75\% of the closest significant peaks are located there, but about 25\% of the time the closest peak is near to point B.  About 1\% of the time, a peak is not found near point B or point C: instead the identified peak is around point A or the nearest noise peak.

\begin{figure}
\begin{center}
\includegraphics[width=9cm]{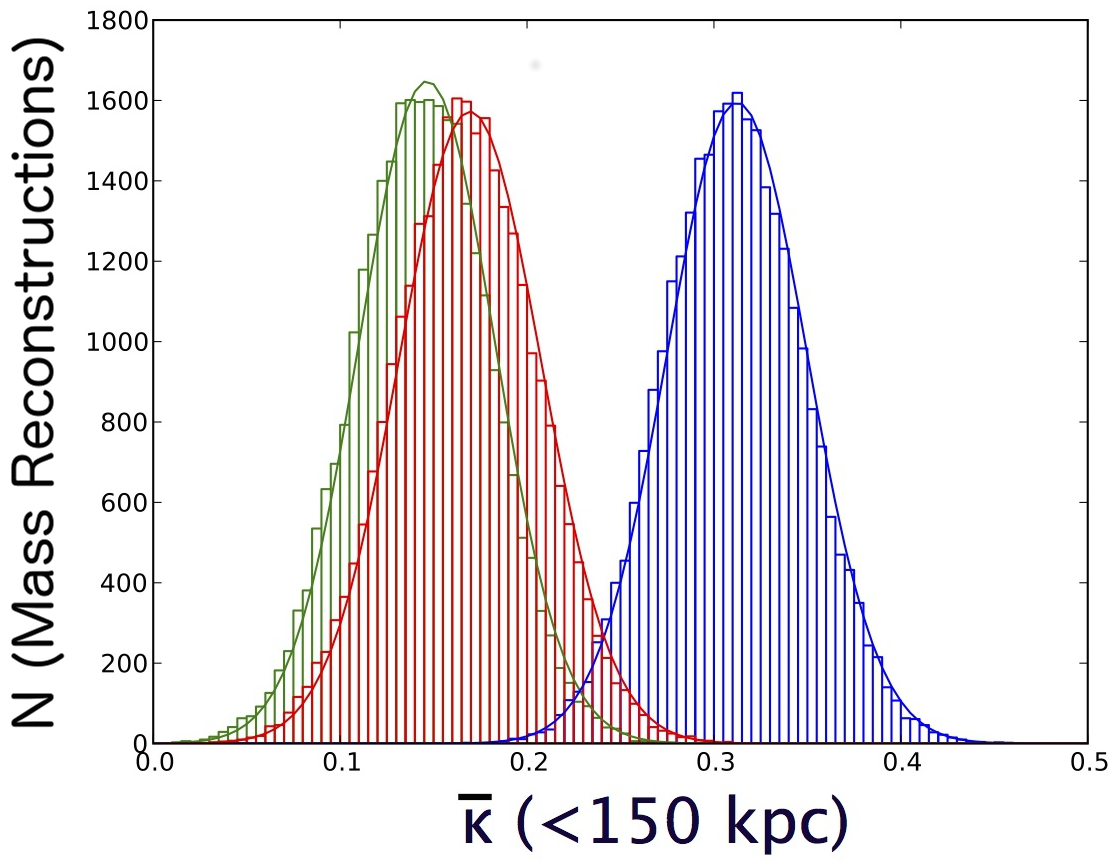}
\caption{For 30,000 mass reconstructions made using bootstrapped resampled catalogues of the selected background galaxies, 
histograms of the number of mass reconstructions, N (Mass Reconstructions), as a function of the mean convergence inside 150\,kpc. The measurements are around fixed points A, B and C and are plotted in blue for A (right distribution), green for B (left distribution), and red for C (middle distribution), respectively.}
\label{appabar}
\end{center}
\end{figure}

For each of the reconstructed mass maps from the bootstrapped catalogues, the mean convergence $\bar\kappa$ inside a radius of 150\,kpc (about 3 times the smoothing scale) of each fixed point was also calculated and the distribution of these values is shown in Fig.\,\ref{appabar}. Note that the measurements around B and C probe overlapping regions.

\begin{figure}
\begin{center}
\includegraphics[width=8cm]{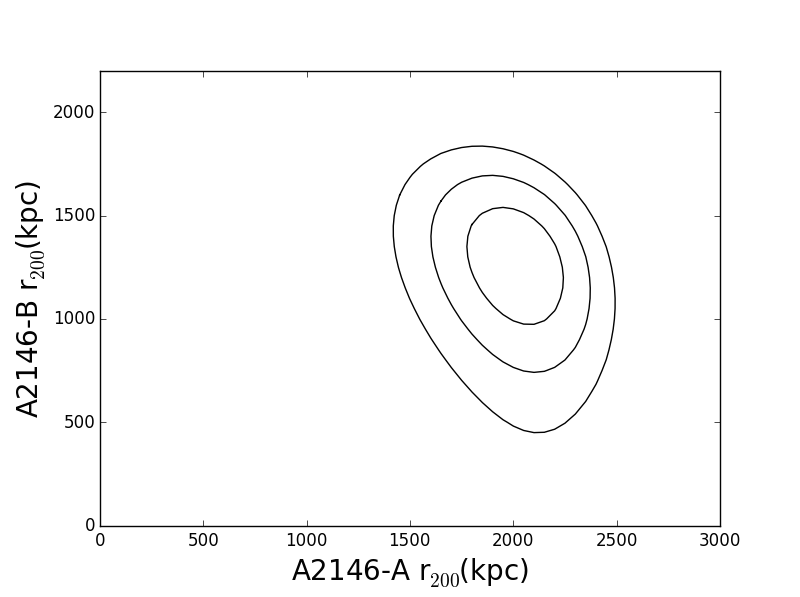}
\caption{Constraints on $r_{200}$ for each of the clusters Abell 2146-A and Abell 2146-B obtained by fitting a 2 component NFW model to the weak lensing data, setting the mass concentration parameter $c=3.5$ for each cluster. The contours are plotted for 1, 2 and 3$\sigma$ confidence intervals.} 
\label{r200}
\end{center}
\end{figure}

In addition to mass reconstruction, we also obtained parameterized mass models consisting of two components. We simultaneously fit two NFW (Navarro, Frenk \& White 1996) mass model components to the 2-D shear estimates obtained from the PSF-corrected ellipticities of the distant galaxies, since each galaxy provides a noisy measurement of the reduced shear. Background galaxies within 100 kpc of the component centres were excluded in order to avoid the break down of the weak lensing regime. The two NFW components were first of all centred on BCG-A and BCG-B. For comparison, we also instead fit models with two components centred on BCG-A and fixed point C, since the reconstructed mass peak in Abell 2146-B appears closer to point C than to BCG-B. Since the bootstrapped resampled mass maps (see statistics in Fig.\,6) some times find a peak closer to BCG-B and some times closer to point C, we also 
centred one component on BCG-A and one component on a grid of locations around the region containing BCG-B and point C in order to find the best fit location for the parameterized model. From the analysis of the bootstrapped resampled catalogues and corresponding mass maps described above, and from strong lensing analysis (Coleman et al. in prep.), the primary mass peak in the system is consistently centred on BCG-A. The expressions for the convergence and shear for the NFW model are given in Bartelmann (1996). Simultaneous fits of the two NFW components each described by a radius $r_{200}$ (or equivalently $M_{200}$) and mass concentration parameter $c$ were carried out. The parameter $r_{200}$ is the radius at which the mean enclosed density of a halo is 200 times the critical density of the universe at the redshift of the halo. The parameter $c = r_{200}/r_{s}$, where $r_{s}$ is the scale radius at which the mass density has an isothermal slope. We fixed $c$ at various values characteristic of galaxy clusters, since there is a well-known degeneracy between $r_{200}$ and $c$ that is exacerbated by the limited spatial extent of the {\it HST} data. The mass of each cluster can then be obtained using
\begin{equation}
M_{200}= \frac{800}{3} \pi r_{200}^3 \rho_{crit}\,;~~~~~~~
\rho_{crit} = \frac{3H^{2}}{8\pi G} \nonumber
\end{equation}
where we take the value of the critical density at $z=0.23$.  Table \ref{NFWtab} shows the best-fit mass models for each cluster component, when $c$ is fixed at various values and when the components are centred on BCG-A and BCG-B. Fig.\,\ref{r200} shows 1, 2 and 3$\sigma$ confidence intervals on the values of $r_{200}$ for the mass model where each cluster component has $c=3.5$. When the two NFW components for the fit are centred on BCG-A and point C instead of on BCG-A and BCG-B, this results in a lower best fit mass for the second NFW component compared with when it was centred on BCG-B. However, fixing one NFW component on BCG-A and using a grid of locations for the second NFW component (in the vicinity of BCG-B and point C), and taking $c=3.5$ for both components, the maximum value of $r_{200}$ for the second component corresponds to a location just on the eastern edge of BCG-B (with a corresponding value of $r_{200}= 1243$ kpc or $M_{200} = 0.27\times 10^{15}M_{\odot}$, about 4\% larger mass than obtained when fixing the component centre to be exactly on BCG-B). We further discuss the parameterized mass models and the mass maps in the Discussion below.

\begin{table}
\caption{Values of $r_{200}$ for best-fit two-component NFW mass models, with fixed values of mass concentration parameter $c$. 
The $\chi^{2}$ values are for the goodness of fit to the selected background galaxies, and the significance is measured from the $\Delta\chi^{2}$ of the best-fit model relative to $r_{200}=0$ (zero mass). The first column indicates the centre of the NFW component (on BCG-A or BCG-B).}
\begin{center}
\begin{tabular}{clllcc}
Cluster&$c$ & $r_{200} (kpc)$ & $M_{200} (10^{15}M_{\odot})$&$\chi^{2}$&significance\\
\hline
A&3.5 & 1971 & 1.09& 0.9607 & 6.7$\sigma$\\
B&3.5 & 1226 & 0.26&0.9607 & 3.5$\sigma$\\\hline
A&4 & 1868 & 0.93& 0.9605 & 6.8$\sigma$\\
B&4 & 1184  & 0.24& 0.9605  & 3.6$\sigma$\\\hline
A&4.5 & 1784 &0.81& 0.9603 & 6.8$\sigma$\\
B&4.5 & 1147 &0.21& 0.9603 & 3.6$\sigma$\\\hline
\end{tabular}
\end{center}
\label{NFWtab}
\end{table}%

\section{Discussion}
\begin{figure*}
\begin{center}
\includegraphics[width=15cm]{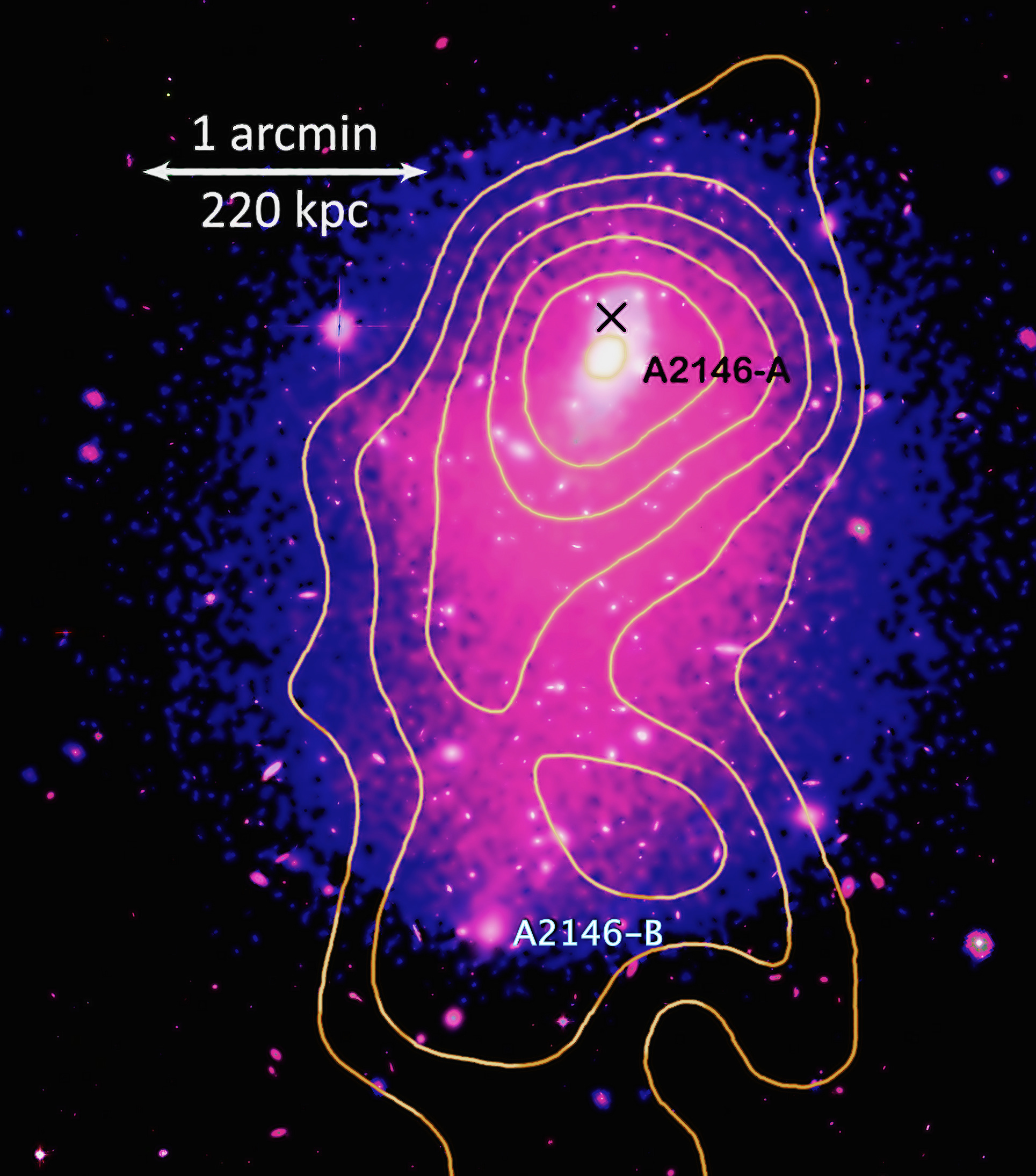}
\caption{Weak lensing mass reconstruction of Abell 2146 (contours) and X-ray intensity from {\it Chandra} X-ray Observatory as described in Russell et al. (2012) in pink and blue shading. The clusters are in the same orientation as in Fig.\,1. The peak of the X-ray emission is marked with an X. The background is a composite from {\it HST} F435W, F606W and F814W observations. }
\label{all}
\end{center}
\end{figure*}

The weak lensing mass map reconstructed from the selected background galaxies (also see Fig.\,4), the {\it Chandra} X-ray map from Russell et al. (2012) and a {\it HST} composite are shown together in Fig.\,\ref{all}. Weak lensing is sensitive to the total mass of a system, irrespective of dynamical state and whether it is luminous or dark matter. Weak lensing mass reconstruction shows a mass peak coincident with the location of the BCG in Abell 2146-A, and a rather more extended mass distribution in Abell 2146-B. 
In weak lensing analysis we need to select galaxies that are background to the system, and without redshift information for the bulk of galaxies we use cuts in colour and brightness in the F814W, F606W, and F435W data. The locations of the peaks and general appearance of the mass maps are consistent for different galaxy selections that we considered during the analysis process. Including galaxies that are in fact foreground or cluster members is an additional source of noise since they are not weakly lensed by the cluster system. However, including galaxies which are background to the system increases the signal to noise of the weak lensing measurement.

Bootstrap resampling of the galaxies selected for the weak lensing analysis was used to explore errors on the mass reconstruction. This procedure preserves the responses of galaxies to the shear field of the system, essentially removing some of the locations at which the shear is sampled and sampling other locations more than once. Fig.\,\ref{diffex} shows two representative examples of mass maps reconstructed from bootstrap resampled catalogues. In order to assess the impact of noise in the mass reconstruction, the statistical properties of the mass maps made from the resampled catalogues can be determined. In particular, Fig.\,6 shows confidence contours on the locations of peaks closest to the BCG in Abell 2146-A and to the BCG in Abell 2146-B and to a point C to the east of Abell 2146-B. This point C was also chosen since a significant peak appeared on many of the bootstrap resampled mass maps. The mass peak closest to the BCG in Abell 2146-A was found to be consistently coincident with the BCG position in the resampled maps. This indicates that the peak of the total mass, dominated by dark matter, near to this BCG is offset from the location of the peak of the plasma seen on the {\it Chandra} X-ray maps. However, the spatial resolution of weak lensing mass maps depends on the scale over which the ellipticities of galaxies must be averaged in order to obtain a significant detection of features above the noise arising from their intrinsic ellipticity dispersion. The smoothing scale was 14.4 arcsec ($\approx 50$\,kpc at the redshift of the cluster system), so the peak of the mass reconstruction is also consistent with the location of the X-ray cool core. This can most easily be seen by referring to the region around the BCG in Abell 2146-A in Fig.\,\ref{boot}, and noting the percentage of bootstrapped mass reconstructions contained within a certain distance from the BCG. At the offset between the X-ray cool core and the BCG (36 kpc), about 68\% of bootstrapped mass reconstructions lie at a closer distance to the BCG. Thus, it is important to note that although the weak lensing mass map is consistent with an offset between the peak in total mass and the X-ray peak, given the errors of the mass reconstruction determined by bootstrap resampling, we can not conclude from weak lensing alone that the peak in total mass and the X-ray peak are offset. 

Fortunately, there are a number of strongly lensed galaxies seen on {\it HST} images that we have used to obtain a higher resolution view of the inner regions of the cluster, and the strong lensing analysis shows an offset between the mass centroid in Abell 2146-A and the X-ray cool core. This study will be presented in Coleman et al. (in prep.).

The mass around Abell 2146-B is rather more extended, with the resampled mass maps sometimes peaking near to the BCG and sometimes to the east of the BCG. About 1\% of the time there is no significant mass peak detected in the vicinity of Abell 2146-B, with the closest peak being either a noise peak or a peak associated with Abell 2146-A. The stability of the reconstructed mass peak coincident with the BCG in Abell 2146-A, and the extended distribution of mass around Abell 2146-B in the 30,000 bootstrap realisations is reflected in Fig.\,7, where the histograms of the mean convergence inside 150\,kpc of the three fixed points (BCGs in Abell 2146-A and Abell 2146-B, and point C to east of BCG in Abell 2146-B) are shown. The histogram of the mean convergence around the BCG in Abell 2146-B has a tail to very low values, consistent with the highest peak in Abell 2146-B appearing offset from the BCG in many of the resampled mass maps. The histogram of the mean convergence around point C is very similar to that around the BCG in Abell 2146-B, indicating a more extended mass distribution in this region. Since the separation of B and C is less than 150\,kpc, the measurements of the mean convergence around these points are not independent.

The larger scale distribution of total mass on the weak lensing mass map appears somewhat more elongated than the hot plasma mapped using {\it Chandra} (Russell et al. 2010; Russell et al. 2012), but both are extended roughly in the direction of the merger axis. As expected, the core passage has displaced gas perpendicular to the merger axis (Russell et al. 2010), consistent with hydrodynamic simulations of merger systems (e.g. Tormen, Moscardini \& Yoshida (2004); Poole et al. 2006). The map of the SZ signal from Rodr\'{i}guez-Gonz\'alvez et al. (2011) is in effect sensitive to the distribution of gas on larger scales and at lower resolution than {\it Chandra}, and it shows an elongation approximately perpendicular to the weak lensing and X-ray maps. As noted by Rodr\'{i}guez-Gonz\'alvez et al. (2011), the peak in the SZ map (in the vicinity of Abell 2146-B) and the peak in the X-ray flux (at the position of the X-ray cool core in Abell 2146-A) are significantly offset, which together with the elongation directions of the X-ray and SZ features are indicative of non-uniformities in gas temperature and pressure consistent with a merger. 

The mass in each cluster component can also be quantified by simultaneously fitting two NFW mass models to the corrected distant galaxy ellipticities. 
With the NFW components centred on the BCGs in Abell 2146-A and Abell 2146-B respectively, the results are shown in Table \ref{NFWtab}. Since the extent of the WFC field of view is 202 arcsec, which corresponds to $\approx$ 750\,kpc at the redshift of the system, it is highly unlikely that the data extend out to the virial radii of the clusters. In weak lensing observations, there is a characteristic strong degeneracy between $r_{200}$ (or $M_{200}$) and $c$. To first order we are sensitive to the mean shear on a data field, and this degeneracy is such that a more massive, lower $c$ cluster can be found to give the same mean shear as a less massive, higher $c$ cluster. This is apparent in Table \ref{NFWtab} where fixing lower values of $c$ results in larger fit values of $r_{200}$ or $M_{200}$. The mass is about 20\% lower on fixing the concentrations of both components to $c=4.5$ rather than $c=3.5$. 
Fixing the concentrations of Abell 2146-A and -B to be equal and simultaneously determining the best fit masses always results in Abell 2146-A being more massive. For best fit models where the concentrations of the two components are allowed to be unequal, and where $c=3.5$ for Abell 2146-B,  Abell 2146-B is more massive only when the concentration of Abell 2146-A is very extreme at $c > 9$. Since this concentration would indeed be unphysical for a galaxy cluster, being more typical of a galaxy, this strengthens the conclusion that Abell 2146-A is more massive than Abell 2146-B. The mass in Abell 2146-B appears extended on the mass map and bootstrapped mass maps; some times the peak is closer to BCG-B and some times to point C. If instead the two NFW components in a parameterized fit are centred on Abell 2146-A and point C, the mass of the second component is lower. Further, allowing the location of the second NFW component to vary over a grid of positions surrounding BCG-B and point C yields a maximum mass for the Abell 2146-B component when the model component centre is just to the east of BCG-B. This location gives a component mass that is about 4\% larger compared with a component centred 
exactly on BCG-B.

As discussed in Russell et al. (2012) there is evidence from the Chandra maps that Abell 2146-B has been disrupted during the merger, consistent with a plume of gas orthogonal to the merger axis which may be the remnant of an X-ray cool core in the cluster. An open question in cluster astrophysics is the degree to which major mergers can disrupt or destroy dense cluster cool cores (eg. G{\'o}mez et al. 2002; Poole et al. 2006; Burns et al. 2008), and hence explain the existence of populations of clusters with and without cool cores. Burns et al. (2008) noted that what may matter most in general for cool core destruction is when in the history of the cluster a merger happens: cool cores seem to be more robust against destruction during mergers occurring long after their formation, and rather more fragile when in their nascent stage. Prior to the start of the merger, Abell 2146-B may originally have been less concentrated than Abell 2146-A, and clusters with lower mass concentrations are more easily disrupted during mergers (Mastropietro \& Burkert 2008). This might explain why Abell 2146-A appears to be the faster moving lower mass bullet from the higher Mach number derived from X-ray maps in Russell et al. (2012), but it presents a higher significance peak in the weak lensing mass map, since the mass around Abell 2146-B is more extended and disrupted by the merger.

We are carrying out simulations to better understand the observed distribution of dark and luminous mass in the system, the pre-collision cluster mass distributions and merger dynamics. Currently, we are applying the Monte Carlo method of Dawson (2013), using constraints from the work here, as well as the results of other observational analyses. In order to understand the origin of the X-ray, SZ, galaxy kinematics and lensing observables, and the seeming physical inconsistency of the X-ray and lensing measurements with regard to X-ray data indicating that Abell 2146-B is the more massive cluster, hydrodynamic simulations of the merger are also being undertaken, constrained by the wealth of observational maps as in Lage \& Farrar (2014) for the Bullet Cluster. 

\section{Conclusions}
We have carried out a weak lensing analysis of the cluster merger system Abell 2146 using data from ACS/WFC on {\it HST}. This is a unique system in that it presents two large shocks on {\it Chandra} X-ray maps (Russell et al. 2010; Russell et al. 2012) and along with dynamical analysis (Canning et al. 2012; White et al. 2015) these limit the time since first core passage to about 0.15 Gyr ago, with a merger axis inclined at about $15^{\circ}$ to the plane of the sky. Thus, we are observing this system at a relatively early stage in the merger, with the clusters still moving apart.

Our weak lensing mass map and parametric models simultaneously fitting two NFW mass components are consistent with Abell 2146-A being the more massive cluster. The mass ratio between the components centred on the BCG in Abell 2146-A and on the BCG in Abell 2146-B is $\approx 3-4:1$ and the total mass is $\approx$1.2$\times$10$^{15}$M$_{\odot}$, assuming NFW mass profiles for each cluster and a mass concentration parameter of $c=4$ for each. Abell 2146-B has a more extended mass distribution, perhaps due to it being a less concentrated cluster that was more disrupted during the merger, and the distribution of mass is likely better described by the mass map rather than by a parametric model. The similarity in the masses of the clusters is in accord with the presence of two large shocks on X-ray maps, in contrast to for example the Bullet Cluster (Markevitch et al. 2004; Clowe et al. 2004; Brada\v{c} et al. 2006; Clowe et al. 2006).

Weak lensing reveals the larger-scale mass distribution, and the peak of the total mass in Abell 2146 is coincident with the BCG in Abell 2146-A. Both the BCG and total mass peak appear to lag the X-ray cool core, though as noted in the Discussion any offset is within the error bar on the mass peak position on the weak lensing map. This error bar was estimated by bootstrap resampling the galaxy catalogues used for the weak lensing analysis, and determining the statistics of the peak locations on 30,000 resampled mass maps. The resolution of weak lensing mass maps is primarily limited by having to average over a sufficient number of galaxies, which have an intrinsic distribution of shapes and orientations, in order to measure a weak lensing signal. A strong lensing analysis of the system, using newly discovered multiple image systems as constraints and revealing the mass in the centre of Abell 2146-A at higher resolution, will shortly be presented in Coleman et al. (in prep.).

Simulations are now underway to understand the merger dynamics and the time-evolution of the luminous and dark matter in the system, constrained by gravitational lensing, X-ray, galaxy kinematics and SZ observables. These simulations will allow us to explore the factors that lead to Abell 2146-A being the more massive cluster from weak lensing, yet the less massive cluster from the analysis of the X-ray data.

\section{Acknowledgements}
Support for program \# 12871 was provided by NASA through a grant from the Space Telescope Science Institute, which is operated by the Association of 
Universities for Research in Astronomy, Inc., under NASA contract NAS 5-26555, in part supporting LJK, DIC, JEC and JAW. 
LJK, JEC, JAW and BEL also acknowledge support from the University of Texas at Dallas. JEC thanks NASA and the Texas Space Grant Consortium for a Graduate Fellowship, and JAW thanks NASA and the 
Texas Space Grant Consortium for a Columbia Crew Memorial Undergraduate Scholarship.  REAC acknowledges support from a scholarship from the Cambridge Philosophical Society and a Royal Astronomical Society grant. 
We thank George Hall and Michael Kesden for discussions.
This research has made use of the NASA/IPAC Extragalactic Database (NED) which is operated by the Jet Propulsion Laboratory, California Institute of Technology, under 
contract with the National Aeronautics and Space Administration.

\appendix

\end{document}